\def\year{2020}\relax
\documentclass[letterpaper]{article} 
\usepackage{aaai20}  
\usepackage{times}  
\usepackage{helvet} 
\usepackage{courier}  
\usepackage[hyphens]{url}  
\usepackage{graphicx} 
\usepackage{graphicx}  
\usepackage{amsmath}
\usepackage{bm}
\usepackage{mathtools}
\usepackage{multirow}
\usepackage{amsfonts}
\newcommand{\citep}{\cite}
\newcommand{\V}[1]{{\bm{\mathbf{\MakeLowercase{#1}}}}} 
\newcommand{\Vhat}[1]{{\bm{\hat \mathbf{\MakeLowercase{#1}}}}} 
\newcommand{\M}[1]{{\bm{\mathbf{\MakeUppercase{#1}}}}} 
\newcommand{\Mbar}[1]{{\bm{\bar \mathbf{\MakeUppercase{#1}}}}} 
\newtheorem{theorem}{Theorem}[section]

\newtheorem{lemma}[theorem]{Lemma}

\newtheorem{proposition}{Proposition}
\newtheorem{remark}{Remark}[section]
\title{A New Framework for Online Testing of Heterogeneous Treatment Effect}
\author{Miao Yu, Wenbin Lu, Rui Song\\ 
Department of Statistics\\ 
North Carolina State University\\
myu12@ncsu.edu, wlu4@ncsu.edu, rsong@ncsu.edu 
}
\begin{document}

\maketitle

\begin{abstract}
We propose a new framework for online testing of heterogeneous treatment effects. The proposed test, named sequential score test (SST), is able to control type I error under continuous monitoring and detect multi-dimensional heterogeneous treatment effects. We provide an online $p$-value calculation for SST, making it convenient for continuous monitoring, and extend our tests to online multiple testing settings by controlling the false discovery rate. We examine the empirical performance of the proposed tests and compare them with a state-of-art online test, named mSPRT using simulations and a real data. The results show that our proposed test controls type I error at any time, has higher detection power and allows quick inference on online A/B testing. 
\end{abstract}

\section{Introduction}
\noindent Randomized controlled experiment, also known as \textit{A/B testing}, is widely used in web facing industry to improve products and technologies in a data-driven manner \citep{kohavi2009controlled}. Most of A/B tests are conducted by performing a formal \textit{null hypothesis statistical testing} (NHST) with the typical \textit{null hypothesis} $H_0: \beta \vcentcolon=\mu_B-\mu_A=0$ to determine if the difference of the metric across two variants is significant or not. The result of a NHST is summarized in a \textit{p-value} and the case that the p-value is less than a preset \textit{significance level} $\alpha$ will lead the null hypothesis to be rejected. A valid testing is able to get a high power to detect the difference if there is, while controlling the \textit{type I error}, i.e., the probability of erroneously rejecting $H_0$, to be less than $\alpha$. 

However, the validity of NHST requires that the sample size is fixed in advance, which is often violated in practice. In A/B testing practice, a fast-paced product evolution pushes its shareholders to continuously monitor the $p$-values and draw conclusions prematurely. In fact, stopping experiments in an adaptive manner can favorably bias getting significant results and lead to very high false positive probabilities, well in excess of the nominal significance level \citep{goodson2014most,simmons2011false}. As an extreme example in \citep{pekelis2015new}, it can be shown that stopping the first time that the $p$-value is less than $\alpha$ actually has type I error probability of 1. Yet for all that, this "peeking" behavior is not without reasons. The time cost and opportunity cost for \textit{fixed-horizon} hypothesis testing are large \citep{ju2019sequential}, so users want to find true effects and stop the experiments as quickly as possible. Moreover, the sample size calculation requires an estimate of the \textit{minimum detectable effect} (MDE). Most users lack good prior knowledge of the trade-off between high detection ability and short waiting time and may want to adjust them after peeking early at results.

Another problem of A/B testing is that it assumes there is only an \textit{average treatment effect} (ATE) in the population of experiment. But underlying this average effect may be substantial variation in how particular subgroups respond to treatments: there may be \textit{heterogeneous treatment effects} (HTE) \citep{grimmer2017estimating}. It might be that the population average effect of a product with a new feature is not significant, but the feature does benefit a lot among particular subgroups of users. In this case, we won't be able to detect those effects and will lose the chance of making profits by promoting new products to those target sub-populations, if only ATE is tested in A/B testing.

To address the continuously monitoring problem, \textit{sequential testing} (ST) was first developed by Wald \citep{wald1945sequential}, who introduced the \textit{sequential probability ratio test} (SPRT). 
ST allows intermediate checks of significance while providing type I error control at any time; see \citep{lai2001sequential} for a survey on sequential testing. Moreover, ST could help decision makers conclude an experiment earlier with often much fewer samples than the fixed-horizon testing \citep{wald1945sequential}. \textit{Mixture sequential probability ratio test} (mSPRT) \citep{robbins1970statistical} and \textit{maximized sequential probability test} (MaxSPRT) \citep{kulldorff2011maximized} are two variants of sequential testing that generalized SPRT to a composite hypothesis. Due to the merits of mSPRT that it is a test with power 1 \citep{robbins1974expected} and almost optimal \citep{pollak1978optimality} with respect to expected time to stop, it was brought to A/B testing by Johari et al. \shortcite{johari2015always,johari2017peeking}. They also proposed a notion of \textit{always valid p-value process} (sequential $p$-values) in the same papers and used it as a tool for converting fixed-horizon \textit{multiple testing} procedures to a sequential version. Later, Malek et al. \shortcite{malek2017sequential} also showed that if the original multiple testing procedure has a type I error guarantee in a certain family (including false discovery rate and family-wise error rate), then the sequential conversion inherits an analogous guarantee.

However, current online testing procedures, such as mSPRT, are not suitable for testing heterogeneous treatment effects due to two aspects. First, they can not accommodate the nuisance parameters in the baseline effects. Second, they may not be able to control the type I error and may lack of power for detecting heterogeneous treatment effects. 
In this paper, we propose a new framework for online testing of heterogeneous treatment effects. The proposed test, named SST, is based on the ratio of asymptotic score statistic distributions, which is able to test multi-dimensional parameters. Furthermore, the asymptotic normality of the score functions guarantees an explicit form of the integral, which allows the integration for the ratio to be efficient. 
At last, we generalize our framework to online multiple testing, which is often the case in industrial practice.

The remainder of this paper is structured as follows. In Section 2, we introduce some preliminary knowledge about fixed-horizon testing and sequential testing. In Section 3, we present the proposed new framework for online testing of heterogeneous treatment effects. We extend SST to multiple testing settings in Section 4 and conduct experiments in Section 5 to compare our framework with the widely-used mSPRT. Finally, in Section 6, we conclude the paper and present future directions.

\section{Preliminaries}
\subsection{Fixed-horizon testing}
\textit{Fixed-horizon testing} is the most widely used procedure in industry where the sample size is fixed in advance. It can be broken down into several steps \citep{lehmann2006testing}:

\textit{Step 1: Determine a desired significance level $\alpha$, minimum detectable effect (MDE) and power at MDE}.
It means that the probability to detect the MDE is at least at the value of power, while the probability of rejecting $H_0$, if it is actually true, is at most $\alpha$.

\textit{Step 2: Calculate/Estimate the minimum sample size $n$}. The sample size $n$ needs to be large enough to achieve the desired power at MDE while controlling type I error at a significance level, but too large sample size will lead to more opportunity cost of waiting for more samples. One need to trade off between these two aspects when choosing sample size.

\textit{Step 3: Collect $n$ samples and compute the observed value of an appropriate test statistics $\Lambda_n$}. The most common test statistics for two-sample tests are z-tests and t-tests, which assume that data are from a normal distribution with known or unknown variance, respectively.

\textit{Step 4: Compute a p-value $p_n$ and reject the null hypothesis if $p_n \leq \alpha$}.  \textit{P-value} is a random variable to denote the probability of seeing a test statistic as extreme as the observed statistics $\Lambda_n$ under null hypothesis, and can be formally defined as 
\begin{equation}
    p_n=\inf \{\alpha: \Lambda_n \geq k(\alpha)\},
    \label{eq1}
\end{equation}
 where $k(\alpha)$ is a critical value depending on significance level and the distribution of $\Lambda_n$ under $H_0$. The critical value is determined such that, under the null hypothesis $H_0$, the event $\Lambda_n \geq k(\alpha)$ occurs with probability no greater than $\alpha$. Since the $p$-value was computed assuming a fixed sample size $n$, we refer to this as a \textit{fixed-horizon p-value}.
Small p-values suggest evidence in support of alternative hypothesis.

A \textit{decision rule} is a pair $(T, \delta)$ representing a testing, where $T$ is a stopping time indicating the sample size at which the test is ended, and $\delta$ is a binary indicator for rejection decision. With the definition of fixed-horizon p-value in (\ref{eq1}), it is obvious to see that $(n, \delta_1)$ and $(n, \delta_2)$ with $\delta_1=1\{p_n \leq \alpha \}$ and $\delta_2=1\{\Lambda_n \geq k(\alpha) \}$ are two equivalent decision rules for fixed-horizon testing. That means the decision rule and p-value can be obtained from each other: find p-value from decision rule $(n, \delta_2)$ by (\ref{eq1}), or make the decision $(n, \delta_1)$ from p-value. Hence, we can actually stop at step 3 and reject $H_0$ if $\Lambda_n \geq k(\alpha)$ for some predetermined significance level $\alpha$. Nonetheless, the decision-making process using p-values is remarkably simple and transparent: one can choose their own significance level and make a valid decision. 

\subsection{Sequential testing}
\textit{Sequential testing}, contrast to fixed-horizon, is a procedure where the decision of terminating the process at any stage of the experiment depends on the results of the observations previously made. It has gained recent popularity in online A/B testing \citep{balsubramani2015sequential,johari2015always} due to its flexibility of continuously monitoring and ending the experiment as soon as significant results are observed.

The decision rules for sequential testing is a nested family of $(T(\alpha), \delta(\alpha))$, parameterized by significance level $\alpha$. It has the following two properties \citep{johari2017peeking}: First, 
the type I error is controlled, that is, $P_{H_0}(\delta(\alpha)=1) \leq \alpha$; Second, $T(\alpha)$ is (almost surely) non-increasing in $\alpha$ while $\delta(\alpha)$ is (almost surely) non-decreasing in $\alpha$. In other words, less stringent type I error control allows the test to stop sooner, and is more likely to lead to rejection.

Similar to fixed-horizon testing, a notion of sequential $p$-values was also introduced for sequential testing and named \textit{always valid p-value process} by \citep{johari2017peeking}: \textit{A sequence of fixed-horizon p-values $(p_n)_{n=1}^\infty$ is always valid if it satisfies the property that $\forall s \in [0,1],\, \mathbb{P}_{H_0}(p_T \leq s) \leq s$ for any given (possibly infinite) stopping time $T$}. It allows the user to trade off detection power and sample size dynamically as they see fit while still control type I error. In the same way, the always valid p-values can be derived from the decision rule for a sequential test, and vice versa. For a given sequential test $(T(\alpha), \delta(\alpha))$,
\begin{equation}
p_n=\inf\{\alpha: T(\alpha) \leq n, \delta(\alpha)=1\}
\label{eq2}
\end{equation} 
defines an always valid p-value process. For any always valid p-value process $(p_n)_{n=1}^\infty$, a sequential test is obtained as follows:
\begin{equation}
T(\alpha)=\inf \{n: p_n \leq \alpha\}
\quad \delta(\alpha)= 1\{T(\alpha) < \infty\}.
\end{equation}

The \textit{mixture sequential probability ratio test} (mSPRT) \citep{robbins1970statistical} is a well studied family of sequential tests. Its test statistic based on the first $n$ observations $\Lambda_{n}^{\pi}$ is a mixture of likelihood ratios against the null hypothesis, with the mixture density $\pi(\cdot)$ over the space for target parameter $\beta$. The decision rule for mSPRT is as below:
\begin{equation}
    T(\alpha)=\inf \{n: \Lambda_n^{\pi} \geq \alpha^{-1}\} \label{eq12} \qquad \delta(\alpha)=1(T(\alpha)<\infty).
\end{equation}
It can be shown that the type I error for mSPRT is well controlled at $\alpha$ by a simple application of \textit{optional stopping theorem} \citep{grimmett2001probability}, since the likelihood ratio under $H_0$ is a nonnegative martingale with initial value equal to one and so is the mixture of such likelihood ratios; see \citep{malek2017sequential,pekelis2015new} for a detailed proof.

Johari et al. \shortcite{johari2017peeking}, recently, have brought mSPRT to online A/B tests where testing parameters $\mu_A$, $\mu_B$ are assumed to be the mean of Bernoulli or normal distribution, depending on whether the data is binary or continuous. They modified the original mSPRT to make it applicable to industrial A/B tests based on some approximation techniques, and empirically showed that the new test has high detection performance with type I error control. 

\subsection{Heterogeneous Treatment Effect}
Up to now, all the online A/B tests we have talked about are focusing only on testing the \textit{average treatment effect} (ATE). However, treatment effects are commonly believed to be varying among individuals, and individual treatment effects may differ in magnitude and even have opposite direction. This is called \textit{heterogeneous treatment effect} (HTE). Testing HTE could help us identify sub-populations where treatment shows better performance and allow personalized treatment as well.

To give a better insight of the difference between ATE and HTE testing, let's take the generalized linear model (GLM) for example,
\begin{align}
    &Y_i \overset{ind}{\sim} \text{Exponential Family}(\gamma_i, \phi), \quad i=1, \cdots, n \nonumber\\
   &f_{Y_i}(y_i| \gamma_i, \phi) = \exp\left\{\frac{y_i \gamma_i - b(\gamma_i)}{a_i(\phi)}+c(y_i,\phi)\right\}, \label{eq3}
\end{align}
where $n$ denotes the sample size, $a_i(\cdot)$, $b(\cdot)$ and $c(\cdot, \cdot)$ are known functions, $\gamma_i$ is the \textit{canonical parameter}, and $\phi$ is a typically known \textit{dispersion parameter}. They are related to the mean and variance of the response through: 
\begin{equation}
  \mu_i=\mathbb{E}(Y_i)={b}'(\gamma_i), \qquad  Var(Y_i)=a_i(\phi)\cdot {b}''(\gamma_i).
\end{equation}
A link function $g(\cdot)$ provides the relationship between the linear predictor and the mean of response:
\begin{equation}
    g(\mu_i)=\eta_i.
    \label{eq4}
\end{equation}
where the linear predictor $\eta_i$ has different forms depending on either ATE or HTE setting.
There is always a well-defined \textit{canonical link} derived from the response's density function, which is a link function such that $g(\mu_i)=\gamma_i$. For example, normal distribution has an identity function $g(\mu_i)=\mu_i$ as the canonical link, Bernoulli has a logit link $g(\mu_i)=\log \frac{\mu_i}{1-\mu_i}$ and Poisson has a log link $g(\mu_i)=\log \mu_i$.

HTE and ATE testings have different assumptions about the form of the linear predictor. ATE testing assumes that
\begin{equation}
    \eta_i=\theta+\beta A_i,
    \label{eq5}
\end{equation}
and test $H_0: \beta=\beta_0$, whereas HTE testing assumes that
\begin{equation}
    \eta_i=\V{\theta}^T \M{X}_i+ (\V{\beta}^T \M{X}_i)A_i, 
    \label{eq6}
\end{equation}
and test $H_0: \V{\beta}=\V{\beta}_0$, where $\M{X}_i$ denotes the covariates vector with the first element being 1 indicating the intercept, and $A_i$ denotes the binary treatment. Note that $\V{\beta}$ and $\V{\theta}$ in HTE testing are both vectors since at least one covariate is considered.

In the case of HTE testing, mSPRT does not work well for the following reasons: 
\begin{enumerate}
    \item The test statistic may not have an explicit form if a conjugate prior $\pi(\cdot)$ for likelihood ratio doesn't exist, as is often the case in HTE testing, e.g., logistic regression. As a result, the computation is inefficient to implement in a streaming environment; 
    \item The nuisance parameter $\V{\theta}$ in the likelihood function is unknown. Even though it can be replaced by its estimator, the resulting test statistics is no longer a martingale and hence the type I error cannot be controlled. Johari, Peke-lis, and Walsh \shortcite{johari2015always} used a sufficient statistic for nuisance parameter and applied \textit{central limit theory} to deal with this issue in A/B tests with Bernoulli distribution. However, this technique failed to be extended to HTE setting.
\end{enumerate}
Therefore, we want to develop a valid online test that can deal with heterogeneous treatment effect.

\section{A New Framework of Sequential Testing}
In this section, we propose a new framework of sequential testing, called \textit{Sequential Score Test} (SST), which is able to test heterogeneous treatment effect while accounting for unknown individual effects. This framework is applicable to independent observations from an exponential family, which includes a large set of commonly used distributions.

Instead using integrated likelihood ratios as in mSPRT, we consider the integration of the ratios of asymptotic score statistic distributions under the local alternative against the null hypothesis. The proposed method can naturally handle nuisance parameters in testing HTE. In addition, the asymptotic representation of the score statistics under the local alternative and the null hypotheses (established in Lemma \ref{lemma1}) can lead to a martingale structure under the null similarly as for the integrated likelihood ratio statistics, and the resulting test statistic have a closed form for integration, which facilitates the implementation of the proposed testing procedure.

\subsection{Sequential Score Test}
Suppose we have i.i.d. data $(Y_{i},A_i,\M{X}_{i})$, where $Y$, $A$, $\M{X}$ respectively denote response, binary treatment and $(p+1)$-dimensional covariates vector including an intercept, respectively. We assume that the distribution of $Y_{i}$ conditional on $(A_i,\M{X}_{i})$ is an exponential family defined in (\ref{eq3})-(\ref{eq4}) with $\eta_i$ in the form of (\ref{eq6}), where $\V{\beta}$ and $\V{\theta}$ denote the heterogeneous treatment effect and baseline effect, respectively. We want to test null hypothesis $H_0: \V{\beta}=\V{\beta}_0$ against local alternative $H_1: \V{\beta}=\V{\beta}_0+\frac{\V{\delta}}{\sqrt{n}}$ \quad($\V{\delta}\neq 0$).

To introduce the test statistic of SST, let's start with some notations. For ease of exposition, we suppose that each group has $n$ observations. Let $\M{S}_{n, \V{\beta}}^{(1)}(\V{\theta}, \V{\beta}_0)$ denotes the score function of $\V{\beta}$ for treatment group ($A=1$) under the null hypothesis $H_0: \V{\beta}=\V{\beta}_0$:
\begin{equation}
    \M{S}_{n, \V{\beta}}^{(1)}(\V{\theta}, \V{\beta}_0)=\sum_{i=1}^n \left( \frac{\partial \mu_i^{(1)}(\V{\beta},\V{\theta})}{\partial \V{\beta}^T}\cdot \frac{(Y_i^{(1)}-\mu_i^{(1)}(\V{\beta},\V{\theta}))}{a_i(\phi) \cdot V_i^{(1)}(\V{\beta},\V{\theta})}\right)\bigg |_{\V{\beta}=\V{\beta}_0} 
    \label{eq7}
\end{equation}
and $\M{S}_{n, \V{\theta}}^{(0)}(\V{\theta})$ denotes the score function of $\V{\theta}$ for control group ($A=0$):
\begin{equation}
    \M{S}_{n, \V{\theta}}^{(0)}(\V{\theta})=\sum_{i=1}^n  \frac{\partial \mu_i^{(0)}(\V{\theta})}{\partial \V{\theta}^T}\cdot \frac{(Y_i^{(0)}-\mu_i^{(0)}(\V{\theta}))}{a_i(\phi)\cdot V_i^{(0)}(\V{\theta})},
    \label{eq8}
\end{equation}
where $\mu_i^{(0)}(\V{\theta})= \mathbb{E}(Y_i|A_i=0,\M{X}_i)$, $\mu_i^{(1)}(\V{\beta},\V{\theta})= \mathbb{E}(Y_i|A_i=1,\M{X}_i)$, $a_i(\phi)\cdot V_i^{(0)}(\V{\theta})=Var(Y_i|A_i=0,\M{X}_i)$ and $a_i(\phi)\cdot V_i^{(1)}(\V{\beta},\V{\theta})=Var(Y_i|A_i=1,\M{X}_i)$. For simplicity, let's assume $a_i(\phi)=a(\phi)$ for all $i$ and $a(\phi)$ is known. 

Consider the following estimated average score $\Mbar{S}_n$ for treatment group (A=1) under $H_0: \V{\beta}=\V{\beta}_0$:
\begin{equation}
\Mbar{S}_n \coloneqq \frac{1}{n}\M{S}_{n, \V{\beta}}^{(1)}(\Vhat{\theta}_n, \V{\beta}_0),
\label{eq9}
\end{equation}
where $\Vhat{\theta}_n$ is the \textit{maximum likelihood estimator} of $\V{\theta}$ calculated based on data from the control group ($A=0$). The idea behind SST is to consider  the test statistic as a mixture of asymptotic probability ratios of $\Mbar{S}_n$, instead of the likelihood ratios, under alternative hypothesis to that under null hypothesis. The test statistic $\Tilde{\Lambda}_n^{\pi}$ is defined as below:
\begin{equation}
    \Tilde{\Lambda}_n^{\pi}=\int \frac{\V{\psi}_{\left(\Mbar{I}_n^{(1)}(\Vhat{\theta}_n)(\V{\beta}-\V{\beta}_0),\, \frac{\M{V}_n(\Vhat{\theta}_n)}{n}\right)}(\Mbar{S}_n)}{\V{\psi}_{\left(\V{0} ,\, \frac{\M{V}_n(\Vhat{\theta}_n)}{n}\right)}(\Mbar{S}_n)}  \pi(\V{\beta}) d\V{\beta},
    \label{eq10}
\end{equation}
where 
\begin{itemize}
    \item $\V{\psi}_{(\V{\mu}, \M{\Sigma})}(\cdot) $ denotes the probability density function of multivariate normal distribution with mean $\V{\mu}$ and variance $\M{\Sigma}$
    \item $\M{V}_n(\V{\theta})=\Mbar{I}_n^{(1)}(\V{\theta}) + \Mbar{I}_n^{(1)}(\V{\theta})\left[\Mbar{I}_n^{(0)}(\V{\theta})\right]^{-1}\Mbar{I}_n^{(1)}(\V{\theta})$
    \item $\Mbar{I}_n^{(1)}(\V{\theta})=-\frac{1}{n}
    \frac{\partial \M{S}_{n, \V{\beta}}^{(1)}(\V{\theta}, \V{\beta}_0)}{\partial \V{\theta}}=\frac{1}{n}\sum_{i=1}^n \left[\frac{\frac{\partial \mu_i^{(1)}(\V{\beta},\V{\theta})}{\partial \V{\theta}^T}\cdot \frac{\partial \mu_i^{(1)}(\V{\beta},\V{\theta})}{\partial \V{\beta}}}{a(\phi)\cdot V_i^{(1)}(\V{\beta},\V{\theta})}\right]\bigg|_{\V{\beta}=\V{\beta}_0}$
    \item $\Mbar{I}_n^{(0)}(\V{\theta})=-\frac{1}{n}
    \frac{\partial \M{S}_{n, \V{\theta}}^{(0)}(\V{\theta})}{\partial \V{\theta}}=\frac{1}{n}\sum_{i=1}^n \left[\frac{\frac{\partial \mu_i^{(0)}(\V{\theta})}{\partial \V{\theta}^T}\cdot \frac{\partial \mu_i^{(0)}(\V{\theta})}{\partial \V{\theta}}}{a(\phi)\cdot V_i^{(0)}(\V{\theta})}\right]$
    \item $\pi(\cdot)$ is a "mixture" distribution over the parameter space denoting the distribution of true effects $\V{\beta}$. It is assumed to be positive everywhere. For ease of computation, we often choose $\V{\beta}\sim \M{MVN}(\V{\beta}_0, \tau^2 \M{I})$, where $\M{I}$ denotes $(p+1) \times (p+1)$ identity matrix and $\tau$ is chosen based on historical data
\end{itemize}

Intuitively, large value of $\Tilde{\Lambda}_n^{\pi}$ represents the evidence against $H_0$ in favor of a mixture of alternatives $\V{\beta} \neq \V{\beta}_0$, weighted by $\V{\beta} \sim \pi(\cdot)$.
The decision rule for SST is quite simple and is shown in (\ref{eq12}). That is, given a significance level $\alpha$, the test stops and rejects the null hypothesis at the first time that $\Tilde{\Lambda}_n^{\pi} \geq \alpha^{-1}$; if no such time exists, it accepts the null hypothesis.
\begin{equation}
   T(\alpha)=\inf \{n: \Tilde{\Lambda}_n^{\pi} \geq \alpha^{-1}\} \label{eq12} \quad \delta(\alpha)=1(T(\alpha)<\infty). 
\end{equation}
The corresponding sequential (always valid) p-value at sample size $n$, by definition of (\ref{eq2}), is the reciprocal of the maximum value of $\Tilde{\Lambda}_n^{\pi}$ up to $n$:
\begin{equation}
    p_n=\frac{1}{\text{max}_{m\leq n} \Tilde{\Lambda}_m^{\pi} }.
    \label{eq14}
\end{equation}
It is obvious to see that the online $p$-value is monotonically non-increasing in $n$ and $p_{T(\alpha)}=\alpha$.

\subsection{Validity of SST}
The intuition of $\Tilde{\Lambda}_n^{\pi}$ being the appropriate test statistics comes from representing the mixture of asymptotic probability ratios of $\Mbar{S}_n$. In this section, we will give the asymptotic distribution of $\Mbar{S}_n$ under null hypothesis and local alternative hypothesis, respectively. Meanwhile, we will offer some insights to demonstrate the approximate validity of SST, that is, the type I error is controlled at large sample size.

The following lemma provides the asymptotic distributions of $\Mbar{S}_n$ with proof shown in the supplemental material.
\begin{lemma}
For generalized linear model in (\ref{eq3})-(\ref{eq4})(\ref{eq6}) and $\Mbar{S}_n$ in (\ref{eq9}), define the \textit{information matrix} for each group as below:
\begin{align}
    \M{I}^{(0)}(\V{\theta})&\coloneqq \mathbb{E}_{(\M{X},\M{Y})}\left[\Mbar{I}_n^{(0)}(\V{\theta})\right]=\mathbb{E}_{(\M{X},\M{Y})}\left[\frac{\frac{\partial \mu_1^{(0)}(\V{\theta})}{\partial \V{\theta}^T}\cdot \frac{\partial \mu_1^{(0)}(\V{\theta})}{\partial \V{\theta}}}{a(\phi)\cdot V_1^{(0)}(\V{\theta})}\right]\\
    \M{I}^{(1)}(\V{\theta})&\coloneqq \mathbb{E}_{(\M{X},\M{Y})}\left[\Mbar{I}_n^{(1)}(\V{\theta})\right]= \mathbb{E}_{(\M{X},\M{Y})}\left[\frac{\frac{\partial \mu_1^{(1)}(\V{\beta},\V{\theta})}{\partial \V{\theta}^T}\cdot \frac{\partial \mu_1^{(1)}(\V{\beta},\V{\theta})}{\partial \V{\beta}}}{a(\phi)\cdot V_1^{(1)}(\V{\beta},\V{\theta})}\right]\bigg|_{\V{\beta}=\V{\beta}_0}
\end{align} 
Then, under null hypothesis $H_0: \V{\beta}=\V{\beta}_0$,
\begin{equation}
    \sqrt{n}\Mbar{S}_n \xrightarrow[H_0]{d} \M{MVN}_{p+1}\left(\V{0}, \M{V}(\V{\theta}_0)\right)
\end{equation}
whereas under local alternative $H_1: \V{\beta}=\V{\beta}_0+\frac{\V{\delta}}{\sqrt{n}}$,
\begin{equation}
    \sqrt{n}\left(\Mbar{S}_n-\M{I}^{(1)}(\V{\theta}_0)(\V{\beta}-\V{\beta}_0)\right) \xrightarrow[H_1]{d} \M{MVN}_{p+1}\left(\V{0}, \M{V}(\V{\theta}_0)\right)
\end{equation}
where $\M{V}(\V{\theta})=\M{I}^{(1)}(\V{\theta}) + \M{I}^{(1)}(\V{\theta})\left[\M{I}^{(0)}(\V{\theta})\right]^{-1}\M{I}^{(1)}(\V{\theta})$, and $\V{\theta}_0$ is the true value of the nuisance parameter.
\label{lemma1}
\end{lemma}

By Lemma \ref{lemma1}, the asymptotic probability ratio of $\Mbar{S}_n$ under local alternative $H_1: \V{\beta}=\V{\beta}_0+\frac{\V{\delta}}{\sqrt{n}}$ against under null hypothesis $H_0: \V{\beta}=\V{\beta}_0$ can be represented as:
\begin{align}
    \lambda_n=\frac{\V{\psi}_{\left(\M{I}^{(1)}(\V{\theta}_0)(\V{\beta}-\V{\beta}_0), \frac{\M{V}(\V{\theta}_0)}{n}\right)}(\Mbar{S}_n)}{\V{\psi}_{\left(\V{0} , \frac{\M{V}(\V{\theta}_0)}{n}\right)}(\Mbar{S}_n)}
    \label{eq25}
\end{align}
Different from likelihood ratio, $\lambda_n$ is not an exact martingale, but we can show that the approximate martingale property does hold when the sample size $n$ is large enough. See the following remark for mathematical expression; the proof can be found in the supplemental material.

\begin{remark}
For generalized linear model in (\ref{eq3})-(\ref{eq4})(\ref{eq6}) and $\lambda_n$ defined by (\ref{eq25}), let $\mathcal{F}_n$ denote the filtration that contains historical information as below:
\begin{equation}
   \mathcal{F}_n=\{(\M{X}_i^{(j)},Y_i^{(j)}), i=1,\cdots,n;\, j=0,1\} 
   \label{eq15}
\end{equation}
Then, under the null hypothesis $H_0: \V{\beta}=\V{\beta}_0$,  
$\mathbb{E}\left[\lambda_{n+1} | \mathcal{F}_n\right]$ is approximately equal to $\lambda_n \cdot \exp \left\{o_p(1)\right\}$. 
\label{remark}
\end{remark}

For practical purpose, we usually replace $\lambda_n$ with its following empirical version $\Tilde{\lambda}_n$:
\begin{equation}
     \Tilde{\lambda}_n = \frac{\V{\psi}_{\left(\Mbar{I}_n^{(1)}(\Vhat{\theta}_n)(\V{\beta}-\V{\beta}_0),\, \frac{\M{V}_n(\Vhat{\theta}_n)}{n}\right)}(\Mbar{S}_n)}{\V{\psi}_{\left(\V{0} ,\, \frac{\M{V}_n(\Vhat{\theta}_n)}{n}\right)}(\Mbar{S}_n)}
     \label{eq13}
\end{equation}
which is exactly the main term in the definition (\ref{eq10}) of $\Tilde{\Lambda}_n^{\pi}$. The empirical ratio $\Tilde{\lambda}_n$ shares the same martingale property as $\lambda_n$ when the sample size is large enough.

Similar to mSPRT in Section 2.2, if we can show that the ratio $\Tilde{\lambda}_n$ is a martingale under null hypothesis $H_0: \V{\beta}=\V{\beta}_0$, the type I error control for SST follows immediately by applying \textit{optional stopping theorem} and the fact that \textit{a mixture of a martingale is also a martingale}. Clearly, as a result of asymptotic distribution and empirical replacement, exact martingale cannot be proved for $\Tilde{\lambda}_n$. But with approximate martingale property in Remark \ref{remark}, the decision rule (\ref{eq12}) for SST approximately controls type I error at small $\alpha$ where large sample size is necessary to reject $H_0$. 

\section{Multiple Testing}
The SST framework can also be applied to \textit{multiple testing}, where more than one treatment variation are compared against a baseline variation, or more than one metric are of interest between two variations. The main problem in multiple comparisons is that the probability to find at least one statistically significant effect across a set of tests, even when in fact there is nothing going on, increases with the number of comparisons \citep{hsu1996multiple}.

In fixed-horizon, \textit{Bonferroni correction} \citep{Miller1966simultaneous} and \textit{Benjamini-Hochberg(BH)}  \citep{benjamini1995controlling} are two well-studied methods designed to address this issue. The Bonferroni correction deals with multiple testing by controlling the \textit{family-wise error rate} (FWER): the probability of making at least one false rejections. Although FWER control provides the safest inference, it is too conservative to offer sufficient detection power. Therefore, the BH procedure is proposed to control the \textit{false discovery rate} (FDR): the expected proportion of the rejections that are false. Both these two procedures take as input the vector of the p-values for each comparison and produce a set of rejections.

In sequential test, the always-valid p-value defined in (\ref{eq2}) works as the ordinary p-value in fixed horizon testing. It is trivial to show that Bonferroni or BH procedure applied on a collection of sequential p-values controls FWER or FDR (respectively) in the presence of arbitrary continuous monitoring \citep{johari2017peeking}. The corresponding algorithms for sequential multiple comparisons under SST framework can be summarized in proposition \ref{prop1} and \ref{prop2}.

\begin{proposition}(Bonferroni Correction for SST). 
For arbitrary stopping time $T$, compute the corresponding sequential p-values $(p_T^{i})_{i=1}^m$ by (\ref{eq14}) for $m$ comparisons. Then reject hypotheses $(1), ...(j)$, where $j$ is the maximal such that $p_T^{(j)}\leq \alpha/m$, and $p_T^{(1)}, ..., p_T^{(m)}$ are the p-values arranged in an increasing order.
\label{prop1}
\end{proposition}

\begin{proposition}(Benjamini-Hochberg Procedure for SST).
For arbitrary stopping time $T$, compute the corresponding sequential p-values $(p_T^{i})_{i=1}^m$ by (\ref{eq14}) for $m$ comparisons. Then reject hypotheses $(1), ...(j)$, where $j$ is the maximal such that:
\begin{equation}
    p_T^{(j)} \leq \frac{\alpha j}{m \sum_{r=1}^m 1/r}
    \label{eq16}
\end{equation}
and $p_T^{(1)}, ..., p_T^{(m)}$ are the p-values arranged in an increasing order.
\label{prop2}
\end{proposition}
Note that the term $\sum_{r=1}^m 1/r$ in (\ref{eq16}) accounts for the fact that the p-values may be correlated \citep{benjamini2001control}.

\section{Experiment}
\subsection{Simulation}
In this section, we compare our SST with the widely-used mSPRT for both A/B tests (two-variations tests) and multiple tests on simulation data generated from combinations of three generalized linear models (\ref{eq3})-(\ref{eq4})(\ref{eq6}) and five types of covariates. The significance level $\alpha=0.05$, null value of testing parameter $\V{\beta}_0=(0,0)$ (for 2-dimensional covariates) or $(0,0,0)$ (for 3-dimensional covariates) and true nuisance parameter $\V{\theta}_0=(0,1)$ (2-dimension) or $(0,1,-1)$ (3-dimension) are fixed for all experiments. Each experiment is repeated 1000 times to estimate type I error and power for SST and mSPRT.

\textbf{Generalized linear models}: We choose three generalized linear models to represent response in different applications. For binary outcomes, such as clicks, conversions, etc., we use logistic regression (Bernoulli distribution). For real-valued response like revenue, ordinary linear regression (normal distribution) is a good choice. If the response are non-negative integers, Poisson distribution which corresponds to log regression is appropriate. However, mSPRT didn't provide the form of test statistics for Poisson distribution, so we only gives our SST result for log regression.

\textbf{Covariates generation}: We consider 5 different distributions for 2 or 3-dimensional ($p=1 \,\text{or}\, 2$) covariates. The first dimension is always 1 to indicate the intercept. The other element of 2-dimensional covariates are generated from normal distribution N$(0,1)$, uniform distribution U$[-1,1]$ and Bernoulli distribution Ber$(0.5)$, respectively. The last two elements of 3-dimensional covariates are generated either from a multivariate normal with mean $(\begin{smallmatrix}
0\\0
\end{smallmatrix})$ and variance $(\begin{smallmatrix}
1 & 0.5\\0.5 & 1
\end{smallmatrix})$, or a hybrid distribution with one variable from N$(0,1)$ and the other one from U$[-1,1]$ independently.

In A/B testing, data are generated in batch with batch size 200, and then are assigned equally to control group and treatment group. After each batch, we compute the corresponding test statistic and reject the null hypothesis the first time it exceeds some predetermined threshold. We also set an upper bound $N=10000$, which means that we would accept the null hypothesis if the test statistic does not exceed the threshold before the data are accumulated to $N=10000$ (for each group). We set the true value of HTE $\V{\beta}$ to be 3 vectors with different scales, including the null value $\V{\beta}_0$. For $\V{\beta}=\V{\beta}_0$, we estimate the type I error by computing the rejection ratio among 1000 repeated experiments. For other two vectors, we estimate the power in the same way.

It shows that the sequential score test is able to control type I error (Table \ref{tb1}), and achieve higher detection power (Table \ref{tb2}) than mSPRT if there is heterogeneous treatment effect. We also find that if there exists individual effects on response, that is, $\V{\theta}\neq \V{0}$, mSPRT may not be able to control type I error (Table \ref{tb1}). That is because the model assumption for mSPRT given in formula (\ref{eq5}) cannot handle individual baseline effects (i.e. $\V{\theta}$ in (\ref{eq6})) and possible treatment-covariates interaction effects (i.e. HTE effects described by $\V{\beta}$ in formula (\ref{eq6})). Therefore, the mSPRT test cannot adjust baseline covariates and may lead to incorrect type I errors for testing HTE. For example, when $\V{\theta} \neq \V{0}$ while $\V{\beta} = \V{0}$ in (\ref{eq6}), mSPRT may reject the null hypothesis due to the outcome difference caused by baseline effects, which may lead to inflated type I errors. On the other hand, when $\V{\beta} \neq \V{0}$, the mSPRT may fail to detect the HTE (lose power) or need to wait a long time to reject the null hypothesis since treatment effects may be masked by individual heterogeneity.

Our proposed test also works with high-dimensional covariates. We conduct additional simulations with 21 covariates (p=20) under logistic regression. Except the first dimension (being 1), the last 20 covariates are independently generated from different distributions. Among these 20 covariates, 7 are generated from normal distribution with variance 1 and different means between -0.3 to 0.3, 8 covariates are from uniform distribution with mean 0 and upper limit between 0.3 to 1, and the last 5 covariates are generated from binomial distribution with probability between 0.1 to 0.5. The individual baseline effect $\V{\theta}$ has two non-zero components. The simulation result shows that our SST still can control type I error under the null (type I error is 0.025, which is less than the significance level $\alpha$), and has reasonable power (i.e. when HTE effects $\V{\beta}$ has three non-zero components with the value of 0.2, the power is 0.731; and when $\V{\beta}$ has three non-zero components with the value of 0.3, the power is 1).

In multiple testing, the configurations of the hypotheses involve $m=64$ hypotheses, $\frac{3}{4}m$ true null hypotheses ($\V{\beta}_0=(0,0)$ or $(0,0,0)$) and the remaining $\frac{1}{4}m$ true alternatives being equally placed at $\V{\beta}_0=(-B,B)$ or $(-B,B,-B)$, where $B=0.1,0.2,0.3,0.4$, respectively. For each comparison, we wait until the data are accumulated to $N=10000$ (for each group) and compute the sequential p-value $p_N$ according to (\ref{eq14}). After applying \textit{Benjamini-Hochberg(BH)} procedure, we get the rejections from which we can estimate FDR and \textit{true positive rate} (TPR), also known as \textit{recall}. The TPR, defined as the proportion of correctly rejections in truly alternatives, is a metric for detection power in multiple testing. Same as A/B testing, the results (Table \ref{tb3}) show that SST applied on multiple testing achieves higher TPR than mSPRT while maintaining FDR in control.

\begin{table}[ht]
    \centering
    \resizebox{.95\columnwidth}{!}{

    \begin{tabular}{c|c|c|c|c|c}
    \hline
    GLM & $\V{\beta}_0$ & $\V{\theta}_0$ & Covariates & Type I error (SST)& Type I error (mSPRT)
    \\
    \hline
    \multirow{4}*{Logistic} & \multirow{3}*{(0,0)}& \multirow{3}*{(0,1)}& N(0,1) & 0.017 & 0.001  \\
    ~& ~& ~& U[-1,1] & 0.019& 0.004 \\
    ~& ~& ~& Ber(0.5) & 0.016 & 0.005 \\
    \cline{2-6}
    \multirow{1}*{Regression}&  \multirow{2}*{(0,0,0)} & \multirow{2}*{(0,1,-1)}& MVN & 0.023 & 0.003\\
    ~& ~& ~& N(0,1)+U[-1,1] & 0.026 & 0.002\\
    \hline
    \multirow{4}*{Linear}&\multirow{3}*{(0,0)}& \multirow{3}*{(0,1)}& N(0,1) & 0.001& 0.132  \\
    ~& ~& ~& U[-1,1] & 0.003& 0.026 \\
    ~& ~& ~& Ber(0.5) & 0.005 & 0.021 \\
    \cline{2-6}
    \multirow{1}*{Regression}& \multirow{2}*{(0,0,0)}& \multirow{2}*{(0,1,-1)}& MVN & $<0.001$ & 0.136\\
    ~& ~& ~& N(0,1)+U[-1,1] & $<0.001$ & 0.201\\
    \hline
    \multirow{4}*{Log} &\multirow{3}*{(0,0)}& \multirow{3}*{(0,1)}& N(0,1) & 0.006& \multirow{5}*{NA}  \\
    ~& ~& ~& U[-1,1] & 0.008&  \\
    ~& ~& ~& Ber(0.5) & 0.006&  \\
    \cline{2-5}
    \multirow{1}*{Regression}& \multirow{2}*{(0,0,0)}& \multirow{2}*{(0,1,-1)}& MVN & 0.003 &  \\
    ~& ~&~& N(0,1)+U[-1,1] & 0.004& \\
    \hline

    \end{tabular}}
    \caption{Estimated Type I error for HTE and ATE testing}
    \label{tb1}
\end{table}

\begin{table}[ht]
    \centering
    \resizebox{.95\columnwidth}{!}{
    \begin{tabular}{c|c|c|c|c|c}
    \hline
     GLM & $\V{\beta}_0$ & $\V{\theta}_0$ & Covariates & Power (SST) & Power (mSPRT)
    \\
    \hline
    \multirow{6}*{Logistic}& \multirow{3}*{(-0.12,0.12)}& \multirow{3}*{(0,1)}& N(0,1) & 0.730& 0.356 \\
    ~& ~& ~& U[-1,1] & 0.709 & 0.514 \\
    ~& ~& ~& Ber(0.5) & 0.215 & 0.079\\
    \cline{2-6}
    \multirow{4}*{Regression}& \multirow{3}*{(-0.15,0.15)}& \multirow{3}*{(0,1)}& N(0,1) & 0.956 & 0.655 \\
    ~& ~& ~& U[-1,1] & 0.938 & 0.851 \\
    ~& ~& ~& Ber(0.5) & 0.436 &0.169\\
    \cline{2-6}
    ~& \multirow{2}*{(-0.12,0.12,-0.12)}& \multirow{2}*{(0,1,-1)}& MVN & 0.544 & 0.384\\
    ~& ~& ~& N(0,1)+U[-1,1] & 0.559 & 0.287\\
    \cline{2-6}
    ~& \multirow{2}*{(-0.15,0.15,-0.15)}& \multirow{2}*{(0,1,-1)}& MVN & 0.875 & 0.636\\
    ~& ~& ~& N(0,1)+U[-1,1] & 0.897 & 0.564\\
    \hline
    
    \multirow{6}*{Linear} & \multirow{3}*{(-0.05,0.05)}& \multirow{3}*{(0,1)}& N(0,1) & 0.685 & 0.607\\
    ~& ~& ~& U[-1,1] & 0.419 & 0.535\\
    ~& ~& ~& Ber(0.5) & 0.053 & 0.099\\
    \cline{2-6}
     \multirow{4}*{Regression}& \multirow{3}*{(-0.08,0.08)}& \multirow{3}*{(0,1)}& N(0,1) & 1& 0.943\\
    ~& ~& ~& U[-1,1] & 0.979 & 0.960 \\
    ~& ~& ~& Ber(0.5) & 0.413 & 0.323\\
    \cline{2-6}
    ~& \multirow{2}*{(-0.05,0.05,-0.05)}& \multirow{2}*{(0,1,-1)}& MVN & 0.400 &0.607  \\
    ~& ~& ~& N(0,1)+U[-1,1] & 0.602 & 0.653\\
    \cline{2-6}
    ~& \multirow{2}*{(-0.08,0.08,-0.08)}& \multirow{2}*{(0,1,-1)}& MVN & 0.994 & 0.955\\
    ~& ~& ~& N(0,1)+U[-1,1] & 0.999 & 0.943\\
    \hline
     \multirow{6}*{Log} &\multirow{3}*{(-0.05,0.05)}& \multirow{3}*{(0,1)}& N(0,1) & 0.499&  \multirow{10}*{NA}\\
    ~& ~& ~& U[-1,1] & 0.132&\\
    ~& ~& ~& Ber(0.5)&0.026&  \\
    \cline{2-5}
    \multirow{4}*{Regression}& \multirow{3}*{(-0.08,0.08)}& \multirow{3}*{(0,1)}& N(0,1) & 0.996&  \\
    ~& ~&~& U[-1,1] & 0.758&  \\
    ~& ~&~& Ber(0.5) & 0.282& \\
    \cline{2-5}
    ~& \multirow{2}*{(-0.05,0.05,-0.05)}& \multirow{2}*{(0,1,-1)}& MVN & 0.242 &\\
    ~& ~&~& N(0,1)+U[-1,1] & 0.726& \\
    \cline{2-5}
    ~& \multirow{2}*{(-0.08,0.08,-0.08)}& \multirow{2}*{(0,1,-1)}& MVN & 0.971&\\
    ~& ~&~& N(0,1)+U[-1,1] &1& \\
    \hline

    \end{tabular}}
    \caption{Estimated power for HTE and ATE testing}
    \label{tb2}
\end{table}

\begin{table}[ht]
    \centering
    \resizebox{.95\columnwidth}{!}{
    \begin{tabular}{c|c|c|c|c|c}
    \hline
    GLM & Covariates & FDR (SST) & FDR (mSPRT)  & TPR (SST) & TPR (mSPRT) 
    \\
    \hline
    \multirow{3}*{Logistic}& N(0,1)&  0.0119& 0.0008 & 0.8038 & 0.7191\\
    ~& U[-1,1] & 0.0059 & 0.0009 & 0.7957 & 0.7662 \\
    \multirow{2}*{Regression}& Ber(0.5) & 0.0067 & 0.0009 & 0.6501 & 0.4761\\
    ~& MVN & 0.0114 & 0.0009 & 0.7664 & 0.7171\\
    ~& N(0,1)+U[-1,1] & 0.0148 & 0.0007 & 0.7775& 0.6944\\
    \hline
    \multirow{3}*{Linear}& N(0,1)& 0.0004 &0.1983 & 1 & 0.3787 \\
    ~& U[-1,1] & 0.0009 & 0.0687 & 0.9994 & 0.2868 \\
    \multirow{2}*{Regression}& Ber(0.5) & 0.0011 & 0.3332 & 0.8725& 0.0504\\
    ~& MVN & 0.0024&0.2013 &0.9999&0.3748\\
    ~& N(0,1)+U[-1,1] & 0.0005 &0.2708 &1& 0.4092\\
    \hline
    \multirow{3}*{Log} &N(0,1)& 0.0011 &\multirow{5}*{NA} &1 & \multirow{5}*{NA}\\
    ~&U[-1,1] & 0.0025 && 0.9750& \\
    \multirow{2}*{Regression}&Ber(0.5) & 0.0019 && 0.8396& \\
    ~&MVN & 0.0010&& 0.9997&\\
    ~&N(0,1)+U[-1,1] &0.0011 && 1&\\
    \hline
    \end{tabular}}
    \caption{Estimated FDR and TPR of HTE and ATE testing for multiple testing}
    \label{tb3}
\end{table}

\subsection{Real Data}
We also compare SST with mSPRT on Yahoo dataset which contains user click events on articles over 10 days. Each event has a timestamp, a unique article id, a binary click indicator, and five user features which are between 0 and 1 and sum to 1 for each user (we only use the last four features). We treat each article as different treatment variations, click actions as the binary responses. Our goal is to test if there is any article effects on user click behaviors with (SST) or without (mSPRT) accounting for the user features. 

We first conduct A/A test to show the validity of test on click events with the most popular article (id=109510) on the date May 1st, 2009, by randomly assigning fake treatment indicators to them. Then we conduct A/B test on events with two most popular articles (id=109510 and 109520) on the date May 1st, 2009. With every 200 events (from both articles) coming in a time sequence, we compute the corresponding test statistics. As soon as the statistics exceed the predetermined critical value ($1/\alpha$), we stop and reject the null hypothesis. If all the data are used up, we accept the null hypothesis. The experiment shows that both SST and mSPRT accept the null hypothesis for A/A test, indicating type I errors are well controlled for both tests under the considered hypotheses. For the A/B test, SST needs $n=19600$ events to get rejection conclusion while mSPRT needs $n=67600$. It means that we are able to discover the difference early by accounting for the covariates. We also provide estimated HTE ($\V{\beta}$ in (\ref{eq6})) and ATE (${\beta}$ in (\ref{eq5})) by fitting logistic regression.

\begin{table}[ht]
    \centering
    \resizebox{.95\columnwidth}{!}{
    \begin{tabular}{c|c|c|c}
    \hline
    Control article id & Treatment article id & HTE ($\V{\beta}$)  & ATE ($\beta$)
    \\
    \hline
     109510 & 109520 & (-0.401, -0.091, -0.068, 0.661, -0.178) & -0.179\\
    \hline
    \end{tabular}}
    \caption{Fitted HTE and ATE}
    \label{tb4}
\end{table}

For multiple test, we choose 10 articles and do pairwise comparisons. Hence, there are $m=45$ comparisons in total. We compute $p_T$ for each pair with $T=20000$ from each article and then apply BH procedure. Among 45 pair comparisons, we reject 43 with SST and 23 with mSPRT.

\section{Conclusions}
We propose a new framework of online test based on the probability ratio of score function. It is able to test a multi-dimensional heterogeneous treatment effect while accounting for the unknown individual effect. The asymptotic normality of the score function guarantees an explicit form, greatly improving the computation efficiency. We provide an online p-value for SST and extend the procedure to online multiple testing. We validate our testing procedure by both theoretical proof and empirical results. We also compare it with a state-of-art online test named mSPRT on simulation and real data. The results show that our proposed test controls type I error at any time, has higher detection power and allows quick inference on online A/B testing. 

There is still some interesting work we may do in the future. The decision rule of our test implies that we can only get rejection conclusions unless we wait essentially indefinitely, which is impossible in practice. This necessitates truncating SST at a maximum size and admitting an inclusive result if we ever reach it, which may diminish the power more or less. How to choose the truncating size to trade off between waiting time and power still remains a problem. 

\bibliography{references}
\bibliographystyle{aaai}

\end{document}